\documentclass[preprint,showpacs,preprintnumbers,amsmath,amssymb,longbibliography]{revtex4-2}

\usepackage{amsmath}
\usepackage[normalem]{ulem}
\usepackage{amsfonts}
\usepackage{amssymb}
\usepackage{graphicx}
\usepackage{color}
\usepackage{xcolor}
\usepackage{subcaption}
\usepackage{setspace}
\usepackage{csquotes}
\usepackage[%
    colorlinks=true,
    pdfborder={0 0 0},
    linkcolor=red
]{hyperref}

\begin{document}

\title{Real and Reciprocal Space Characterization of the 3-Dimensional Charge Density Wave in Quasi-1-Dimensional CuTe}

\author{Fei Guo$^{1,2}$}
\author{Michele Puppin$^{1,2}$}
\author{Lukas Hellbr\"{u}ck$^{1,2}$}
\author{Arnaud Magrez$^{1}$}
\author{Eduardo B. Guedes$^{3}$}
\author{Igor Sokolovi\'{c}$^{4}$}
\author{J. Hugo Dil$^{1,2,3}$}

\affiliation{
$^{1}$Institute of Physics, \'{E}cole Polytechnique F\'{e}d\'{e}rale de Lausanne, CH-1015 Lausanne, Switzerland\\ 
$^{2}$Lausanne Centre for Ultrafast Science (LACUS), \'{E}cole Polytechnique F\'{e}d\'{e}rale de Lausanne (EPFL),
CH-1015 Lausanne, Switzerland\\
$^{3}$Photon Science Division, Paul Scherrer Institut, CH-5232 Villigen, Switzerland\\
$^{4}$Institute of Applied Physics, Technische Universit\"{a}t Wien, 1040 Vienna, Austria}

\date{\today}
\begin{abstract}
Low-dimensional materials are susceptible to electronic instabilities such as charge density waves (CDWs), originating from a divergence in the Lindhard electron response function, combined with a finite electron-phonon coupling strength. In this report, we present a detailed characterisation of the CDW in the quasi-one-dimensional material CuTe, including (1) direct visualization of lattice distortion seen with non-contact atomic force microscopy in real space, (2) the out-of-plane momentum dependency of the CDW gap size of the quasi-1-dimensional bands, by angle-resolved photoemission spectroscopy, (3) coherent dynamics of a photoexcited phonon mode seen by time- and angle-resolved photoemission spectroscopy, with frequency and wavevector $\textbf{\textit{q}}_\text{CDW}$ corresponding to the soft phonon modes predicted by theory. Furthermore, we find that the CDW gap closes through a transient band renormalisation. We thus confirm that, despite the quasi-one-dimensional characteristics of CuTe, it hosts inherently 3-dimensional CDWs.
\end{abstract}

\maketitle

\section{Introduction}
Low-dimensional materials are a vibrant field of research due to their rich number of competing broken symmetry ground states \cite{Young:2014,Hedge:2022}, which are promising for novel electronic functionalities \cite{Ahmad:2021, Kim:2023} and the possibility of ultrafast optical manipulation \cite{Slobodeniuk:2023}. An ubiquitous effect is the susceptibility to charge ordering instability, such as charge density waves (CDWs). In the past few decades, there have been great efforts to understand the driving mechanisms of several CDW-hosting systems \cite{Zhu:2015}, for example, quasi-2D transition metal dichalcogenides (TMDCs) \cite{Borisenko:2008,Borisenko:2009}, rare-earth tritellurides ($RTe_3$s) \cite{Sacchetti:2007,Brouet:2008}, as well as in quasi-one-dimensional systems \cite{Tomeljak:2009,Nicholson:2017}. The formation of CDWs can be due to an electronic instability driven by Fermi surface nesting \cite{Whangbo:1991,Laverock:2005}, combined with electron-phonon coupling \cite{Calandra:2011,Luo:2022}. Electron-electron couplings are believed to be the driving force for CDW in different scenarios, for example in excitonic insulators \cite{Jerome:1967}, or via the band Jahn-Teller effect \cite{Hughes:1977}. In reality, CDWs are likely to arise due to an interplay between multiple of the above factors, which cannot be easily individually resolved. The characterization of materials hosting CDWs requires a quantitative evaluation of the direction and periodicity of charge density modulation represented by the CDW wavevector $\textbf{\textit{q}}_\text{CDW}$, along with the energy scale of such modulation in the form of a transition temperature $T_\text{CDW}$, or the magnitude of the energy gap, as well as a case-specific evaluation of the driving many-body couplings.

\begin{figure}
    \centering
    \includegraphics[width=0.9\textwidth]{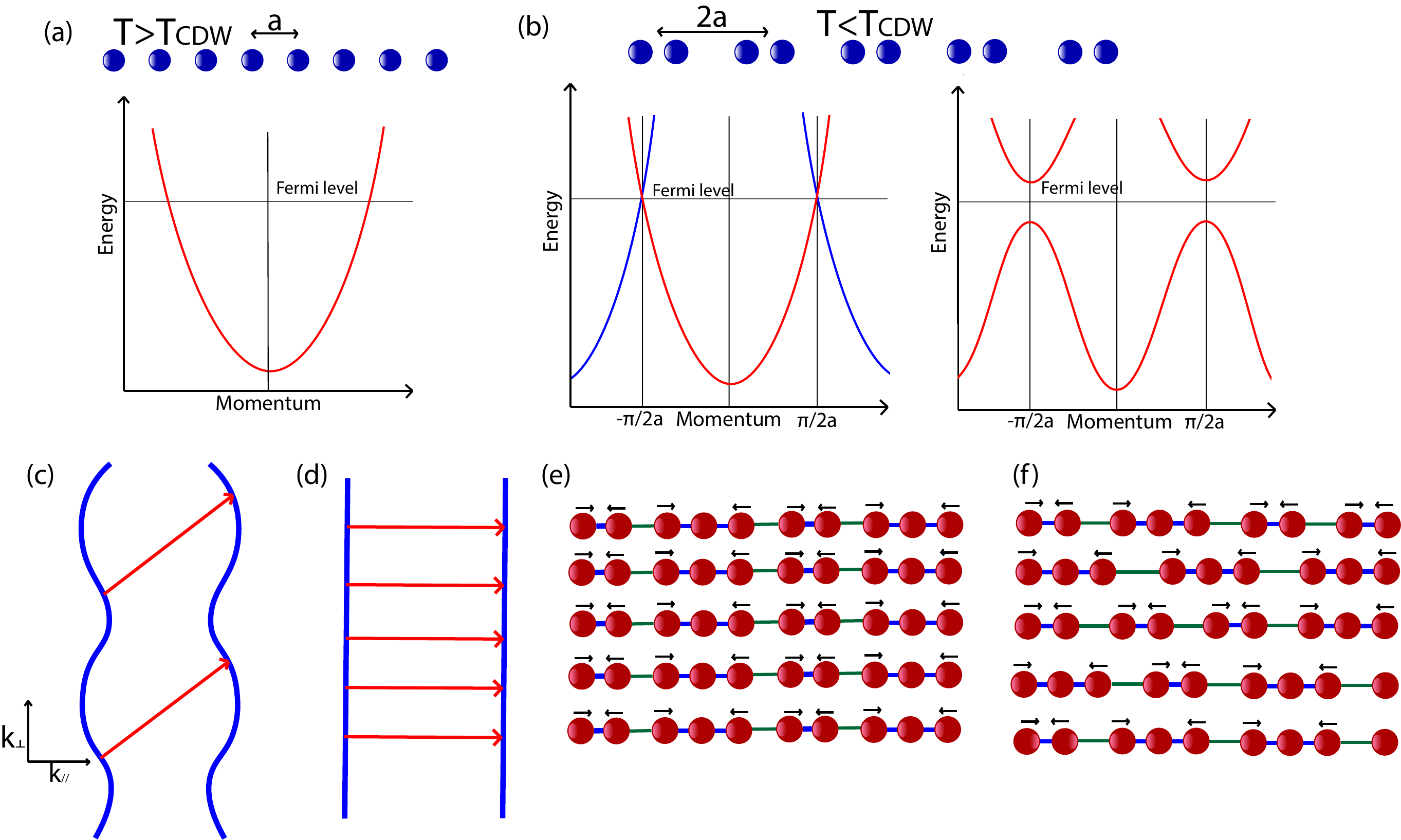}
    \caption{Schematic drawing of (a) undistorted and (b) Peierls distorted lattices and electronic band dispersion. Fermi surface nesting of (c) 3D and (d) 1D CDWs in the first definition in terms of $\textbf{\textit{q}}_\text{CDW}$ \cite{Strocov:2012}, with Fermi surface (blue) and nesting wavevector $\textbf{\textit{q}}_\text{CDW}$ (red). Lattice distortions of (e) 3D and (f) 1D CDWs in the second definition in terms of inter-layer collectivity \cite{Nicholson:2017}.}
    \label{schematics2}
\end{figure}

In the simplest model, the formation of a CDW in quasi-1D materials follows a Peierls scenario, in which pairs of atoms dimerize. This modulation overcomes Coulomb repulsion by opening up a gap at the Fermi energy and hence lowers the total energy in the system. The formation of this superstructure introduces an additional periodicity and `replica' bands, as shown in Fig.\ref{schematics2}(a) and (b). In this model, the Lindhard response function $\chi(\textbf{\textit{q}})$ diverges whenever $\epsilon_k=\epsilon_{k+q}$, i.e. two parts of the Fermi surface are connected by a wavevector $\textbf{\textit{q}}$, a condition referred to as Fermi surface nesting. However, this divergence is not protected with respect to carrier scattering and small geometrical deviations from perfect nesting conditions \cite{Johannes:2008}. Given this fact, one might expect that CDWs could only survive in quasi-1D systems with nearly perfect nesting, but there is plenty of evidence that CDWs exist in 2D \cite{Neto:2001, Tsen:2015, Hossain:2017} and even 3D \cite{Kim:2021, Liang:2021} systems. Therefore, Fermi Surface nesting and electron-phonon coupling are certainly not the only mechanism of CDW formation for higher-dimensional materials.

The dimensionality of CDWs in low-dimensional materials is widely discussed in the literature, but not uniquely defined. For instance, some authors define a 3-dimensional CDW as one where the CDW-associated lattice distortion is not purely in-plane, and, as shown in Fig.\ref{schematics2}(c), there is a finite out-of-plane component of the CDW wavevector $\textbf{\textit{q}}_\text{CDW}$ \cite{Strocov:2012}. Thus a 1-dimensional CDW in this picture would correspond to a 1D Fermi surface with a nesting wavevector $\textbf{\textit{q}}_\text{CDW}$ purely along one in-plane direction as in Fig.\ref{schematics2}(d). Other authors define a 3-dimensional CDW where all layers in the material have the same $\textbf{\textit{q}}_\text{CDW}$ and modulate coherently, but $\textbf{\textit{q}}_\text{CDW}$ itself does not need an out-of-plane component \cite{Nicholson:2017}. A 1-dimensional CDWs in this picture would be an individual modulation on each atomic chain without any inter-layer collectivity, despite having the same $\textbf{\textit{q}}_\text{CDW}$. This distinction is shown schematically in Fig.\ref{schematics2}(e) and (f). At this point, it is worth mentioning that a purely 1D system cannot sustain stable long-range orders due to quantum and thermal fluctuations, and inter-layer collectivity is a prerequisite to the formation of charge and spin  density waves \cite{Gruner:1988, Gruner:1994}.

The subject of this article, copper telluride (CuTe), is a quasi-1D material, with the atomic structure shown in Fig.\ref{AFM1}(a) \cite{Stolze:2013}, which has been reported to undergo a CDW phase transition at $335\,$K. The electronic band structure of CuTe has been investigated with ARPES \cite{Zhang:2018}, revealing a momentum-dependent gap as large as $190\,$meV which forms along quasi-1D Fermi surface sheets. The static electronic band structure and the CDW gap were reported as a function of temperature and doping, to visualize the closure of the gap. First-principle calculations of the phonon spectrum were also presented showing soft phonon modes at the observed nesting wavevector, which also corresponds to maxima of the calculated electron scattering susceptibility. These results thus indicate that electron-phonon coupling and Fermi surface nesting indeed contribute dominantly to the formation of the CDW in CuTe. Detailed electronic structure and ultrafast nonequilibrium carrier dynamics near the Brillouin zone center of CuTe has also been studied with low-photon-energy ARPES and trARPES \cite{Zhong:2024}. Moreover, the dimensionality of the CDW in CuTe has been investigated by ultrafast optics \cite{Nguyen:2024}, in which inter-layer coupling has been observed at low temperatures. In this work, we expand on those results and further elucidate the 3-dimensionality of the CDW in CuTe, by providing both real and reciprocal space investigations.

In the following, we present a thorough investigation of CuTe. We first directly visualize the spatial charge ordering modulations via non-contact atomic force microscope (ncAFM), then, we measure with angle-resolved photoemission spectroscopy (ARPES) the electronic band dispersion of CuTe, including the dispersion orthogonal to the surface plane $k_z$. Our data reveals a $k_z$-dependent gap, together with evidence of inter-layer couplings in the ordering of the CDW. Finally we investigate electron-phonon coupling in the system by femtosecond time-resolved ARPES (trARPES), showing coherent oscillation of specific location of the band structure. The oscillation correspond to a soft phonon mode linked to the CDW, but are strongly coupled with inherently 3D states at the Fermi surface. Further, we can follow the ultrafast closing of the CDW gap through a renormalisation of the quasi-1D bands. In light of these results, we discuss the dimensionality of the CDW in CuTe and show that under all typical definitions of dimensionality the CDW in CuTe is 3-dimensional, despite the fact that both the atomic and electronic structures show remarkable quasi-1D features.

\section{Methods}

High quality single crystals of CuTe were synthesized using the flux technique. A mixture of Cu and Te (atomic ratio 1:20) was sealed under vacuum within a quartz ampoule. This ampoule was then placed in a vertical furnace and heated to 600$^\circ$C, where it remained for 12 hours before being gradually cooled to 450$^\circ$C at a cooling rate of 10$^\circ$C per hour, and then further cooled to 325$^\circ$C at a slower rate of 1$^\circ$C per hour. Following the crystal growth process, the resulting crystals in the solidified Te flux were transferred in a quartz tube furnace and maintained at 270$^\circ$C under dynamic vacuum. After the complete removal of the Te flux by sublimation, golden CuTe crystals were obtained. The structure and stoichiometry were confirmed by single-crystal diffraction and X-ray fluorescence spectroscopy.

Non-contact atomic force microscopy (ncAFM) measurements were performed using a Omicron qPlus low-temperature head mounted in an ultrahigh vacuum (UHV) chamber. Stiff qPlus sensors \cite{giessibl2019qplus} ($k=1800$\,N$\cdot$m$^{-1}$, $Q$=5000--30000, $f_0$\,$\in$\,[25-45]\,kHz) with a a sharp W tip \cite{setvin2012ultrasharp} were used for imaging. Cantilever deflection was supported with an in-vacuum cryogenic preamplifier \cite{huber2017low}. Prior to cleaving a CuTe sample and imaging the cleaved surface, sharp W tips were additionally decorated with a sharp Cu pyramid at the apex by deliberate tip restructuring on a Cu(110) surface. Damping of external vibrations was assisted by suspending the whole chamber with 36 bungee cords \cite{schmid2019device}.

Single-crystal CuTe samples were prepared by attaching them to standardized sample plates and attaching a stainless steel top post using conductive silver epoxy. The silver epoxy was cured by baking this structure at $80^\circ$C for 4 hours in ambient air. Samples were cleaved at room temperature in a chamber with an UHV base pressure below $1\times10^{-10}$\,mbar, quickly transferred \textit{in-situ} to a separate chamber with the base pressure below $1\times10^{-11}$\,mbar where they were introduced to the measurement head and subsequently imaged at $T=5\,$K. Such treatment of cleaved surfaces results in a negligible amount of adsorbates, even on much more reactive surfaces \cite{sokolovic2019incipient}.

For ARPES measurements samples were prepared in the same way as for the ncAFM measurement. For the $k_z$ resolved ARPES measurements samples were cleaved \textit{in-situ} in ultrahigh vacuum below $5\times10^{-10}$\,mbar and at $T=24\,$K, and for the trARPES measurements samples were cleaved \textit{in-situ} in ultrahigh vacuum at $T=70\,$K. High-resolution ARPES measurements as a function of photon energy were taken at the SIS-ULTRA beamline of the Swiss Light Source, Paul Sherrer Institut. All band structures were measured with circular $C+$ polarized light. trARPES measurement were taken at the LACUS facility at the EPFL \cite{Crepaldi:2017b}, which has a high-harmonic generation (HHG) source of extreme ultraviolet (XUV) laser, along with an infra-red pump laser of wavelength 780\,nm (photon energy 1.6\,eV), which provides a time resolution better than 100\,fs. The fluence of the pump pulse on sample surface was 1.3\,mJcm$^{-2}$. trARPES data were collected with $p$-polarized HHG light with a photon energy of 40.35\,eV.

\section{Periodic Lattice Distortion: Direct Evidence of CDW in CuTe} \label{AFM}
\begin{figure}
    \centering
    \includegraphics[width=\textwidth]{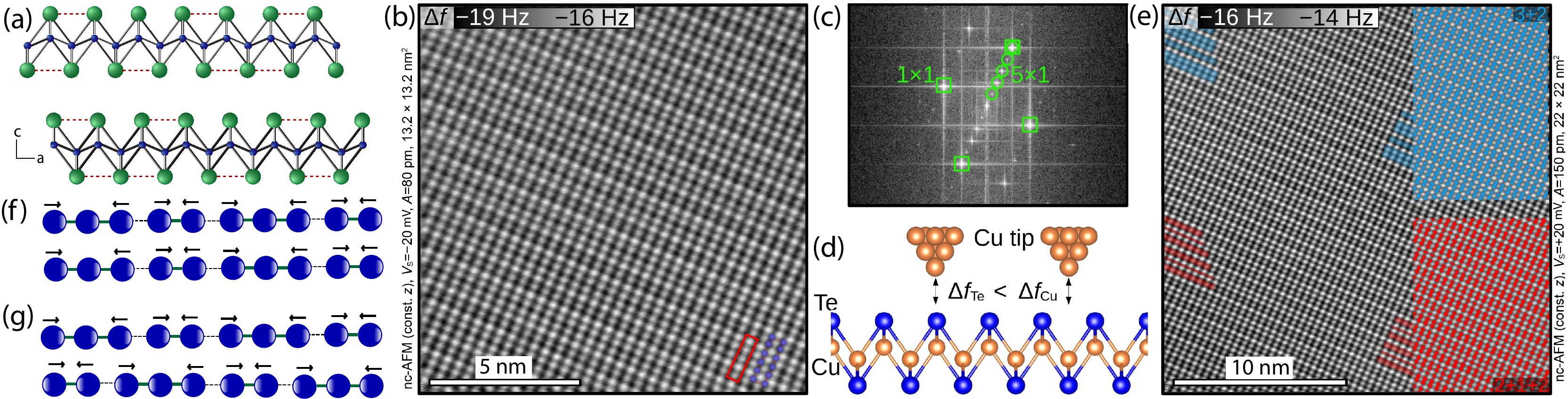}
    \caption{(a) Atomic structure of modulated CuTe, each Cu (blue) is coordinated with 4 Te (green) atoms, 2 or 3 Te atoms group together in the CDW state. (b) Small area ncAFM image: Te atoms are imaged as dark spheres (attraction) while white spheres indicate absence of atoms (only long-range forces). Te atoms are highlighted as blue spheres and the $5\times 1$ repeat unit cell is indicated in red. (c) Fast Fourier transform of ncAFM image, with $5\times1$ reconstruction in one direction and only $1\times1$ peaks in the perpendicular direction. (d) Schematic drawing of the atomic structure of CuTe and ncAFM contrast formation: Surface-terminating Te atoms act with chemical attraction (negative $\delta f$) towards the Cu-terminated tip, while the Cu atoms below the surface primarily contribute with long-range van der Waals forces ($\delta f$ closer to zero). (e) ncAFM image on extended area, with 2 domains of CDWs with periodicities $2+1+2$(red) and $3+2$(blue) respectively. Schematic drawing of two different bilayer arrangements of the distorted atomic chain: (f) 3+2 on 3+2, and (g) 3+2 on 2+3.}
    \label{AFM1}
\end{figure}

The crystal structure of CuTe is shown in Fig.\ref{AFM1}(a), the 1D nature is reflected by the appearance of chains of Cu atoms, each coordinated with 4 Te atoms. Non-contact atomic force microscopy (ncAFM) was used to visualize the real space CDW. A topographic image of a high-quality surface is shown in Fig.\ref{AFM1}(b). The surface is seen to undergo a $5\times1$ lattice modulation, as highlighted by a fast Fourier transform in Fig.\ref{AFM1}(c) which shows a unidirectional $5\times1$ reconstruction. The ncAFM image of a larger area in Fig.\ref{AFM1}(e) shows that the CDW modulation is not a simple $5\times1$, but rather proceeds either as $2+1+2$ [red in Fig.\ref{AFM1}(e)] or $3+2$ [blue in Fig.\ref{AFM1}(e)] modulation, in accordance with a recent STM investigation \cite{Kwon:2024}. On a surface patch of $20\times 20\,\text{nm}^2$ as in Fig.\ref{AFM1}(e) both domains are present without intermixing, in line with the previously assigned wavelength of the Higgs mode to approximately $24\,$nm \cite{Kwon:2024}.

The modulation wavevector of the CDW is $\textbf{\textit{q}}_\text{CDW}=(0.4\textbf{a*},0\textbf{b*},\frac{1}{2}\textbf{c*})$ , in which $\textbf{a*}$ is the reciprocal lattice vector in the chain direction, and $\textbf{c*}$ is the reciprocal lattice vector in the out-of plane direction \cite{Stolze:2013}. The magnitude of the $\textbf{c*}$ component of $\textbf{\textit{q}}_\text{CDW}$ is dictated by the orthorhombic point group symmetry \cite{Janner:1977}. Therefore, in reality we should expect an actual superstructure with a thickness of 2 layers. This can explain the structure of the $2+1+2$ domains seen in Fig.\ref{AFM1}(e). In order for $\textbf{\textit{q}}_\text{CDW}$ to overlap with the reciprocal lattice vectors, it needs to be multiplied by $2$ in the $\textbf{c*}$ direction and $2.5$ in the $\textbf{a*}$ direction. For a commensurate CDW, a superstructure with 2.5 lattice sites is not possible, so the lattice reconstructs with either 2 or 3 sites, alternating to finally form a periodicity of 5 lattice sites. Then, the periodicity of 2 in the $\textbf{c*}$ direction comes into play: the $3+2$ superstructure has broken inversion symmetry, so there are 2 nonequivalent arrangements of 2 layers, $3+2$ on $3+2$, or $3+2$ on $2+3$. These are shown schematically in Fig.\ref{AFM1}(f) and (g). With the $3+2$ on $2+3$ arrangement, the ncAFM tip can detect different modulations on the 2 topmost layers, appearing like a $2+1+2$ modulation.

Given this 3-dimensional superstructure in real space, it is then instructive to explore the electronic band dispersion in 3-dimensional reciprocal space.

\section{Electronic Band Dispersion of CuTe} \label{ULTRA}

\begin{figure}
    \centering
    \includegraphics[width=0.95\textwidth]{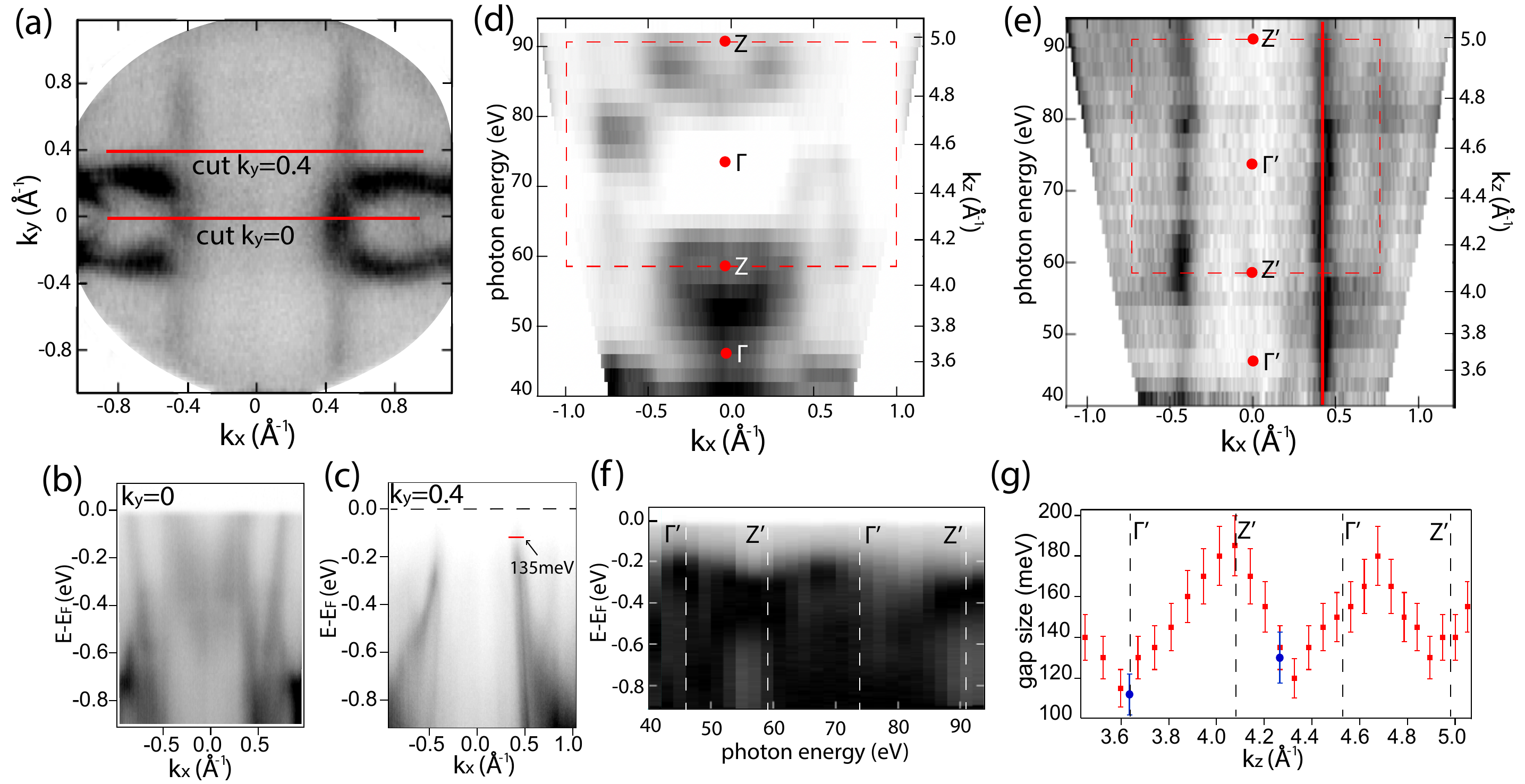}
    \caption{(a) Constant binding energy surface of CuTe taken at $h\nu=65\,$eV with $E-E_F=-0.15$eV, the $k_y$ values at which $h\nu$ scans were performed are marked with red lines. (b) Band dispersion at $k_y=0\,\mathring{A}^{-1}$, with $h\nu=65\,$eV. (c) Band dispersion at $k_y=0.4\,\mathring{A}^{-1}$, with $h\nu=65\,$eV. (d) $h\nu$ dependency at $k_y=0\,\mathring{A}^{-1}$ and $E-E_F=-0.35$eV, with $k_z$ value calculated for normal emission on the right axis. (e) $h\nu$ dependency at $k_y=0.4\,\mathring{A}^{-1}$ and $E-E_F=-0.2$eV, with $k_z$ value calculated for normal emission on the right axis. (f) Energy spectra as a function of photon energy for $k_y=0.4\,\mathring{A}^{-1}$ and $k_x=0.42\,\mathring{A}^{-1}$ marked in (e) by a red solid line. (g) CDW gap size variation with $k_z$ corrected for off-normal emission angles, the 2 data points in large blue circles are extracted from Fermi surface measurements at $h\nu =46\,$eV and $h\nu =65\,$eV}
    \label{FSMz}
\end{figure}

ARPES is the most direct way to visualize the electronic band structure, where the onset of a CDW state is characterized by the opening of a band gap. A constant energy surface of CuTe probed by ARPES is shown in Fig.\ref{FSMz}(a), which shows a central feature extending in the $k_x$ direction, and 2 quasi-1D bands extending in the $k_y$ direction \cite{Zhang:2018}. In the CDW state, these bands display a $k_y$-dependent gap, with a maximum of $190\,$meV at $k_y=0.3\,\mathring{\text{A}}^{-1}$. Here we investigate the $k_z$-dependence of the band structure and gap size, by tuning the photon energy.

We first focus on the cut $k_y=0$, the band structure of which is shown in Fig.\ref{FSMz}(b), and vary the photon energy from $41\,$eV to $93\,$eV, with a step of $2\,$eV. The resulting $k_z$ dispersion is shown in Fig.\ref{FSMz}(d), in which the high symmetry points $\Gamma$ and Z are indicated with the bulk Brillouin zone. It is clear that the bands at $k_y=0$ disperse strongly along the $k_z$ direction and thus have a 3D nature, as also expected from calculations \cite{Stolze:2013}. A similar procedure was carried out for the cut at $k_y=0.4\,\mathring{\text{A}}^{-1}$ across the quasi-1D bands, the dispersion of which is shown in Fig.\ref{FSMz}(c). To stay at a constant $k_y$ value, the tilt angle $\theta$ was adjusted at each photon energy according to the relationship $k_y=\frac{\sqrt{2mE_k}}{\hbar}\sin\theta$. The obtained dispersion is shown in Fig.\ref{FSMz}(e), in which the projected high symmetry points $\Gamma'$ and $\text{Z}'$ are indicated with the bulk Brillouin zone. These bands barely show any dispersion as a function of $k_z$. Combined with the chain-like crystal structure and the fact that there is also little dispersion along $k_y$, these bands show evidently quasi-1D character.

The absence of band dispersion does not mean that also the CDW gap has to remain constant. To explore a possible variation of gap size, the band structure at $k_x=0.42\,\mathring{\text{A}}^{-1}$ is shown in Fig.\ref{FSMz}(f). A periodic variation of the band maximum as a function of $h\nu$ can be clearly observed. The CDW gap size was extracted for each photon energy by tracing the energy distribution of intensity, and the binding energy at which the band intensity decreases by $70\%$ is marked as the gap edge, in accordance with previous work \cite{Zhang:2018, Nguyen:2024}. The gap size is plotted as a function of the calculated $k_z$ in Fig.\ref{FSMz}(g), with 2 extra data points from Fermi surface measurements at $h\nu =46\,$eV and $h\nu =65\,$eV to further confirm the gap size, and with an inner potential of $9.5\,$eV extracted from the $h\nu$ periodicity of the 3D bands in Fig.\ref{FSMz}(d). Whereas the maximum gap size is similar to recent studies \cite{Zhang:2018, Nguyen:2024}, the systematic variation as a function of $h\nu$ has to the best of our knowledge not been studied before. Interestingly, this variation does not follow the periodicity of the Brillouin zone, but the difference in $k_z$ between the 2 gap size maxima in Fig.\ref{FSMz}(g) is $0.60\,\mathring{\text{A}}^{-1}$. This is between the out-of-plane modulation wavevector $0.5\text{c}^*=0.45\,\mathring{\text{A}}^{-1}$ and the out-of-plane reciprocal wavevector $\text{c}^*=0.90\,\mathring{\text{A}}^{-1}$. The physical origin of the observed periodicity, and its relation with the CDW modulation requires further investigation. Nevertheless, this observed periodicity of the CDW gap along the $k_z$ direction further supports the 3D nature of the CDW.

The above results indicate that CuTe hosts a 3-dimensional CDW in terms of the modulation wavevector $\textbf{\textit{q}}_\text{CDW}$, which has a finite component in the z-direction. Also, the prerequisite of inter-layer collectivity to the formation of CDWs, and the absence of Luttinger liquid-like behaviour \cite{Haldane:1981} suggests that all the layers should modulate collectively at low temperatures. This is confirmed by ref.\cite{Nguyen:2024} which shows that at temperatures below $220\,$K, quantum fluctuations are suppressed and all layers modulate collectively in alternate anti-phase. Combining this with our results provides solid evidence that CuTe, albeit quasi-1D, hosts 3-dimensional CDWs with respect to both common definitions.

\section{Coherent Dynamics of CuTe} \label{LACUS}
Electron-phonon coupling is a critical ingredient in the CDW-forming mechanism, here we investigate it by observing the coherent dynamics of the band structure, initiated by a short optical perturbation. Through the oscillation of the electronic bands, one can find the frequencies associated to strongly coupled phonon modes that drives the Fermi surface nesting. For this purpose, pump-probe time-resolved ARPES (trARPES) is a powerful tool, as it can identify directly coupling to specific bands with momentum resolution \cite{Giovannini:2020, Boschini:2024}.

\begin{figure}
    \centering
    \includegraphics[width=\textwidth]{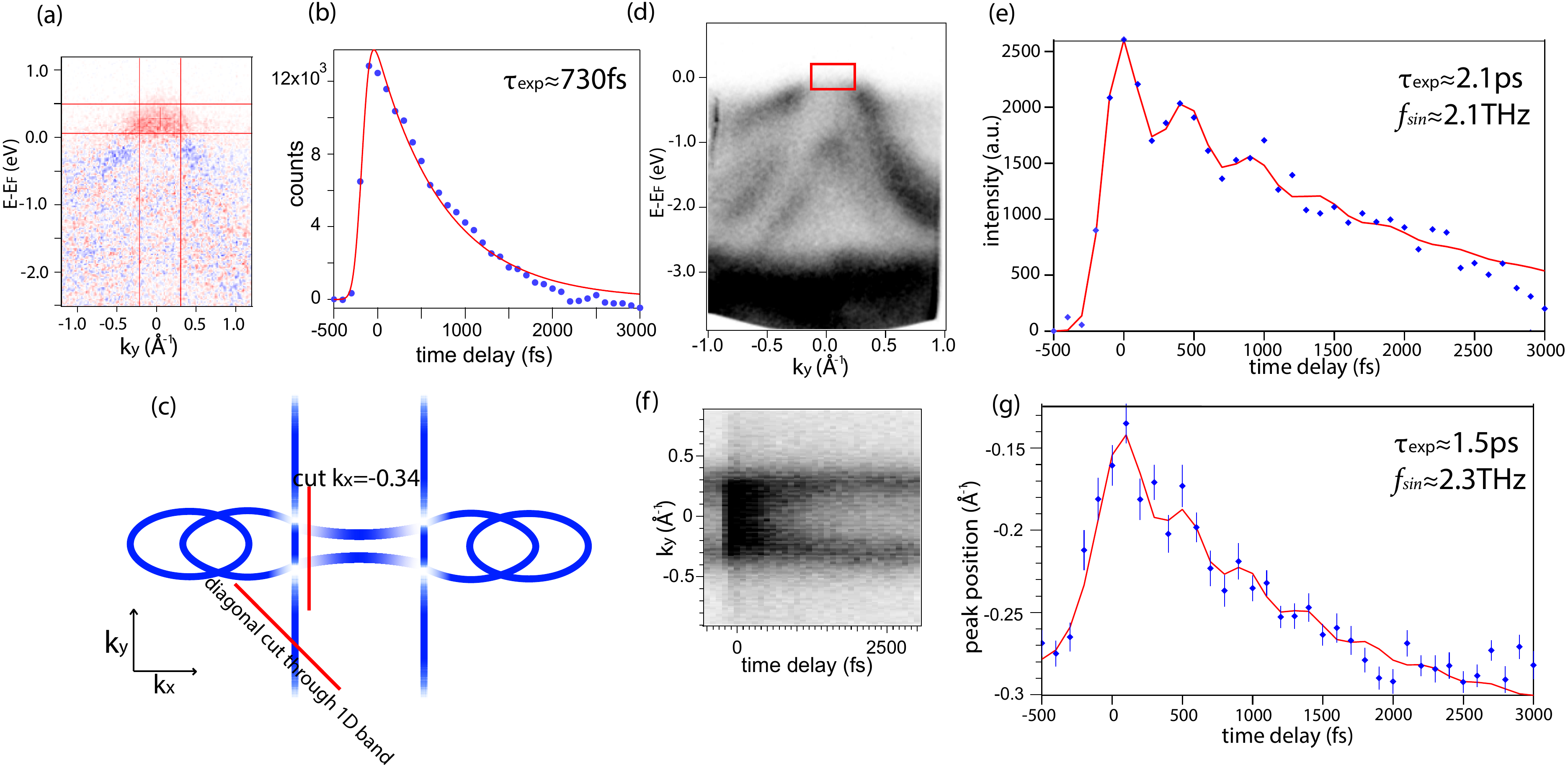}
    \caption{(a) Difference spectrum measured with trARPES at $k_x=-0.34\,\mathring{\text{A}}^{-1}$ for a time delay of 100~fs: red denotes extra intensity compared to negative time delay, and blue denotes vanished intensity compared to negative time delay. (b) Integrated intensity in marked region in (a), as a function of time delay. (c) Schematic drawing of Fermi surface of CuTe. (d) Band map with $k_x=-0.34\,\mathring{\text{A}}^{-1}$ marked in (c). (e) Time-delay distribution of intensity integrated in the area marked in (d), along with line fitting with an exponential multiplied by a sinusoidal function. (f) Momentum distribution of intensity at constant energy $E-E_F=-0.35$eV as a function of time delay. (g) Time-delay distribution of the negative-$k_y$ intensity peak position from (f). All of the above data are obtained at a photon energy of $h\nu = 40.35\,$eV.}
    \label{dynamics}
\end{figure}

To that end, we performed trARPES in the LACUS facility \cite{Crepaldi:2017b}. In this experiment, $1.6\,$eV pump infrared laser pulses were used to excite electrons to states above the Fermi level, then, after an adjustable time delay, an extreme ultraviolet (XUV) laser of $40.35\,$eV photon energy was used to probe the band structure by photoemission. Fig.\ref{dynamics}(a) demonstrates a time delay difference spectrum measured by this technique: red denotes positive signal, corresponding to an increased intensity at a time delay of 100~fs compared to negative time delays (probe before the pump), and blue denotes an intensity decrease. In Fig.\ref{dynamics}(a), the red area with $E-E_F>0$ indicates excited states above the Fermi level. Since the bands cross the Fermi level, the excited states above the Fermi level relax quickly following an exponential decay with $\tau\approx730\,$fs, as shown in Fig.\ref{dynamics}(b).

We first focus on the bands at $k_x=-0.34\,\mathring{\text{A}}^{-1}$, as marked on the schematic Fermi surface in Fig.\ref{dynamics}(c). According to Fig.\ref{FSMz}(b), these bands show strong 3D characteristics. The corresponding band map, and area of interest around Fermi level, are shown in Fig.\ref{dynamics}(d). To track the phonon dynamics, the intensity evolution in this area as a function of delay time is shown in Fig.\ref{dynamics}(e). Clear intensity oscillations can be observed, and can be fitted with an exponential decay with a decay time of $2.1\pm0.1\,$ps and oscillation frequency $f\approx2.1\pm0.1\,$THz, which corresponds to $69\pm3\,\text{cm}^{-1}$. Another way to observe these oscillations is to track directly the movement of bands as a function of pump-probe time delay. Fig.\ref{dynamics}(f) shows the momentum distribution of intensity at a constant binding energy $E-E_F=-0.35\,$eV with varying time delays. The band maximum at negative $k_y$ was extracted as a function of time delay in Fig.\ref{dynamics}(g), on which similar oscillations were fitted with a decay time of $1.5\pm0.1\,$ps and oscillation frequency of $f\approx2.3\pm0.2\,$THz, which corresponds to a phonon frequency of $77\pm7\,\text{cm}^{-1}$. The above findings indicate that with the pump laser, we have photo-excited a phonon mode of CuTe. Interestingly, \textit{ab-initio} calculations of the phonon spectrum \cite{Zhang:2018} indicate softening for two phonon modes at $\textbf{\textit{q}}=(0.4,0,0)$ and $\textbf{\textit{q}}=(0.4,0,0.5)$ respectively, with softened phonon frequencies in the range of $60-85\,\text{cm}^{-1}$, comparable with the observed phonon frequency. Moreover, 2 collective phonon modes have been observed in CuTe by ultrafast optics with frequencies centered at $1.64\,$Hz and $2.26\,$Hz respectively \cite{Nguyen:2024}. With our trARPES measurements on the 3D states, however, we only observe the latter phonon mode.

\begin{figure}
    \centering
    \includegraphics[width=\textwidth]{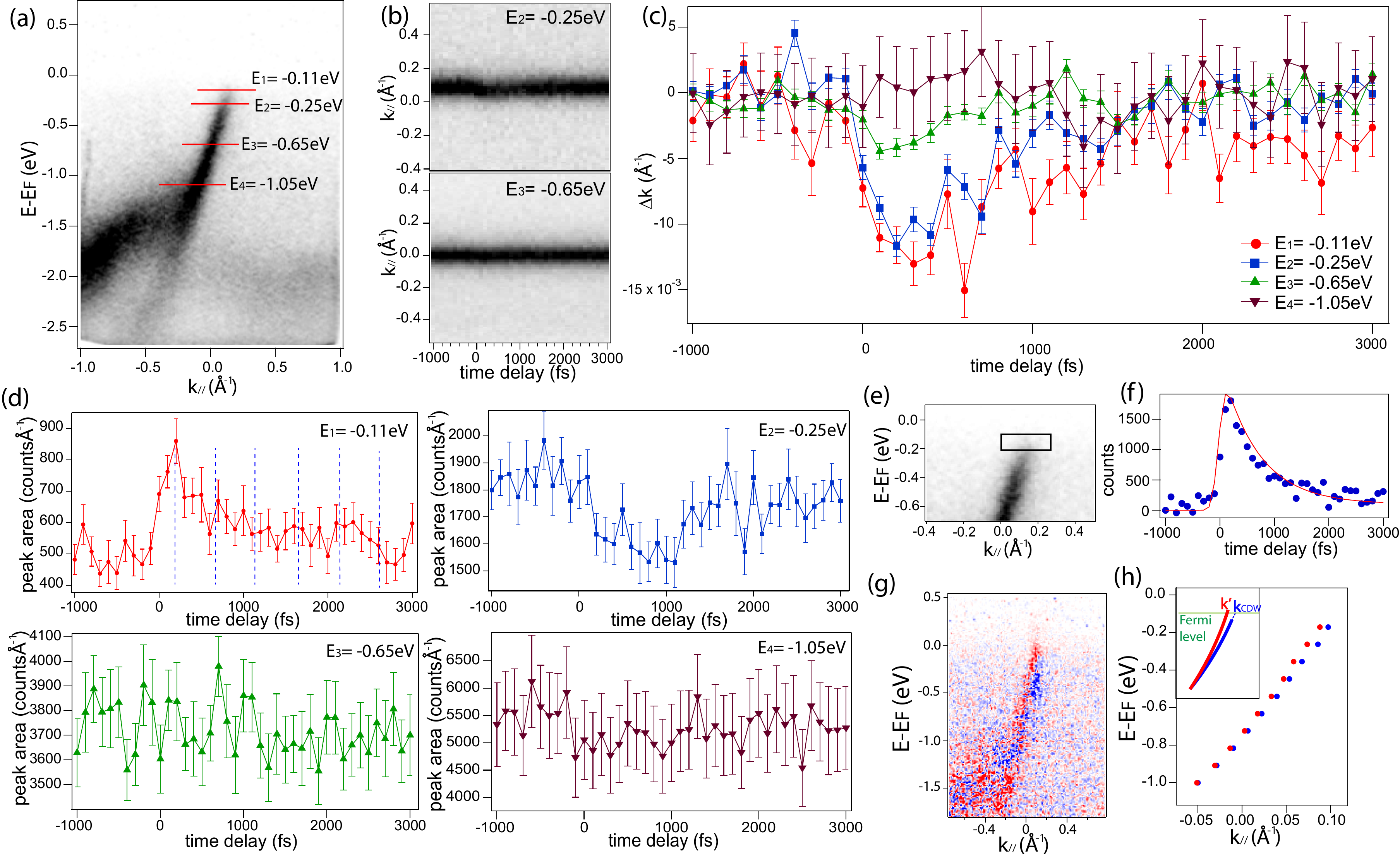}
    \caption{(a) Band map with diagonal $k_{//}$ marked in Fig.\ref{dynamics}(c). (b) Representative momentum distributions of intensity as a function of time delay, at binding energies $E_2$ and $E_3$ marked in (a). (c) Peak position as compared to negative time delays, as a function of time delay, at all binding energies marked in (a). (d) Peak area as a function of time delay, at binding energies marked in (a); blue dashed lines indicate the time period extracted from the coherent oscillations in Fig.\ref{dynamics}(e). (e) Zoom-in from (a) with integration region indicated. (f) Integrated intensity in the boxed region in (e) as a function of time delay, with a fitted relaxation time of $690\,$fs. (g) Difference spectrum of quasi-1D band at a time delay of $200\,$fs, as compared to negative time delays. (h) Band position before (blue) and $200\,$fs after (red) optical pumping plotted by fitting MDCs at different binding energies, inset: schematic drawing of band renormalization. All of the above are taken at a photon energy of $h\nu = 40.35\,$eV.}
    \label{1d_dynamics}
\end{figure}

Gap-filling dynamics has been studied extensively in CDW materials, and is often accompanied by an oscillation of in-gap intensity or gap size \cite{Shen:2008, Maklar:2021, Maklar:2022, Rettig:2016}. To study the behaviour of the bands around the CDW gap, trARPES measurements of the 1D band were taken in a diagonal $k_x-k_y$ direction, around the 1D band as shown by the diagonal red line in Fig.\ref{dynamics}(c), and the corresponding band map is shown in Fig.\ref{1d_dynamics}(a). For each of the marked binding energies in Fig.\ref{1d_dynamics}(a),the momentum distribution of intensity is visualized as a function of time, as representatively shown for $E_2=-0.25\,$eV and $E_3=-0.65\,$eV in Fig.\ref{1d_dynamics}(b). From these momentum distributions we extracted the temporal evolution of the fitted peak position with respect to negative time delays, as shown in Fig.\ref{1d_dynamics}(c). Furthermore, to characterize the intensity evolution, we calculated the peak area (= FWHM peak width$\times$peak intensity) for each binding energy, as shown in Fig.\ref{1d_dynamics}(d). Furthermore, the integrated intensity in the boxed area in Fig.\ref{1d_dynamics}(e) has been plotted as a function of time delay in Fig.\ref{1d_dynamics}(f) for comparison. At binding energy $E_1=-0.11\,$eV, which is within the CDW gap, we see a pronounced renormalization in band position and a gap filling which decays on the order of $700\,$fs, essentially identical to the decay time scale of $\tau \approx 690\,$fs obtained from the integrated intensity in Fig.\ref{1d_dynamics}(e). However, even with the oscillation periodicity of the 3D bands from Fig.\ref{dynamics}(e) superimposed, it is difficult to identify any clear oscillations within the experimental uncertainty. At binding energy $E_2=-0.25\,$eV, a comparable band renormalization is observed, along with an intensity depletion which recovers with a slightly longer time scale. At the 2 higher binding energies, we see no pronounced band renormalization or intensity change within the experimental signal to noise ratio. 

The phenomenon of band renormalization upon optical excitation is expected to arise from transiently enhanced screening by high-energy electrons and holes. It has been reported for gapless states of the nodal-line semimetal ZrSiSe \cite{Gatti:2020}, and for CDW-state HoTe$_3$ \cite{Rettig:2016}. For the latter, enhanced screening transiently reduces the tight binding parameter $t_{\perp}$, and nesting conditions are improved as a result of this renormalization.  On CuTe, in contrast, this screening-induced renormalization has a consequence of transient CDW melting. Fig.\ref{1d_dynamics}(g) shows a difference spectrum taken at a time delay of $200\,$fs, in which we see a renormalization of the band gradient. As a result, the momentum at which the band crosses, or would cross, the Fermi level changes by approximately $0.015\,\mathring{\text{A}}^{-1}$. This situation is shown explicitly in Fig.\ref{1d_dynamics}(h). The main figure shows the peak positions obtained by fitting MDCs at different binding energies for negative time delay (blue) and a time delay of  $200\,$fs (red), while the inset schematically illustrates the observed band renormalization. Since the renormalized band (red) crosses the Fermi level with a momentum which slightly differs from the nesting wavevector, the nesting condition is no longer fulfilled and the CDW is transiently melted.


Overall, with our laser pump-probe trARPES experiment, we discover that, the photoresponse of the system to femtosecond excitation comprises of a coherent response to the softened phonon mode, which appear to be strongly coupled to the 3D states, and an incoherent hot electron response lasting several hundreds of femtoseconds. For the 1D states, only an incoherent dynamics was observed and no coherent oscillations could be extracted with confidence. This dynamics is associated to an ultrafast gap melting and dispersion renormalization towards deteriorated nesting conditions, where the maximum renormalization occurs at around $200\,$fs, shorter than a single oscillation period observed on the 3D states. However, due to the 3D nature of the CDW in CuTe, we cannot rule out the possibility that coherent oscillation on the 1D states becomes observable at other photon energies.

\section{Conclusion}

In this work, we combined different techniques and demonstrated for CuTe, that the real space CDW superstructure and the reciprocal space CDW gap size both have a periodicity in the out-of-plane direction. These are evidences that the CDW wavevector $\textbf{\textit{q}}_\text{CDW}$ has a finite out-of-plane component. Together with the prerequisite of inter-layer collectivity for CDW formation which is indeed observed below $220\,$K \cite{Nguyen:2024}, we confirm that the CDW in this quasi-1D material is 3D by all definitions. Our results reveal that interlayer coupling effects play an important role in the CDW of CuTe, and highlight the importance of combining multiple experimental probes.

\section*{Acknowledgement}

The authors thank Ulrike Diebold at TU Wien for supporting the ncAFM measurements. F.G. and J.H.D. acknowledge support from the Swiss National Science Foundation (SNSF) Project No. 200021-200362.

\bibliography{References_SOIS}

\end{document}